\documentclass[%
 aip,
 pof,%
 amsmath,amssymb,
preprint,%
]{revtex4-1}

\usepackage{graphicx}
\usepackage{xcolor}
\usepackage{natbib}
\usepackage{amsmath}
\usepackage{amssymb}
\usepackage{bm}
\usepackage{ar}

\newcommand{\ie}{i.e.,\ }
\newcommand{\eg}{e.g.,\ }

\definecolor{cinnamon}{rgb}{0.82, 0.41, 0.12}

\newcommand{\secref}[1]{\mbox{section \ref{#1}}}

\newcommand{\tabref}[1]{\mbox{table \ref{#1}}}

\newcommand{\equref}[1]{\mbox{equation (\ref{#1})}}

\newcommand{\figrefC}[2][]{\mbox{Figure \ref{#2} (#1)}}
\newcommand{\figref}[2][]{\mbox{figure \ref{#2} (#1)}}
\newcommand{\figrefSC}[1]{\mbox{Figure \ref{#1}}}
\newcommand{\figrefS}[1]{\mbox{figure \ref{#1}}}

\begin{document}

\title{Low Reynolds number turbulent flows over elastic walls}
\author{Marco E. Rosti}
\email[Corresponding author: ]{marco.rosti@oist.jp}
\affiliation{Complex Fluids and Flows Unit, Okinawa Institute of Science and Technology Graduate University, 1919-1 Tancha, Onna-son, Okinawa 904-0495, Japan}
\author{Luca Brandt}
\affiliation{Linn\'{e} Flow Centre and SeRC (Swedish e-Science Research Centre), KTH Department of Engineering Mechanics, SE 100 44 Stockholm, Sweden}

\begin{abstract}
We study the laminar and turbulent channel flow over a viscous hyper-elastic wall and show that it is possible to sustain an unsteady chaotic turbulent-like flow at any Reynolds number by properly choosing the wall elastic modulus. We propose a physical explanation for this effect by evaluating the shear stress and the turbulent kinetic energy budget in the fluid and elastic layer. We vary the bulk Reynolds number from 2800 to 10 and identify two distinct mechanisms for turbulence production. At moderate and high Reynolds numbers, turbulent fluctuations activate the wall oscillations, which, in turn, amplify the turbulent Reynolds stresses in the fluid. At very low Reynolds number, the only production term is due to the energy input from the elastic wall, which increases with the wall elasticity. This mechanism may be exploited to passively enhance mixing in microfluidic devices.
\end{abstract}

\maketitle

\section{Introduction} \label{sec:introduction}
In typical microfluidic applications, the Reynolds number is very small and the flow is laminar. If chaotic mixing is not induced by the device geometry via fully three-dimensional flow fields, mixing is due to molecular diffusion only, resulting in long diffusion times, which limits the efficiency of micro-scale devices. In this context, we study here the feasibility to use a soft elastic wall to enhance mixing by inducing self-sustained chaotic velocity fluctuations also at very low Reynolds numbers.

Several strategies have been proposed in the past to increase mixing in micro-devices, which can be classified into passive and active: in the former, the mixing is enhanced through curved streamlines \citep{knight_vishwanath_brody_austin_1998a, bessoth_manz_others_1999a,  stroock_dertinger_ajdari_mezic_stone_whitesides_2002a, lee_chang_wang_fu_2011a, lee_wang_liu_fu_2016a}, while in the latter the flow is made unsteady by an external actuation \citep{glasgow_aubry_2003a, bazant_squires_2004a, mensing_pearce_graham_beebe_2004a, kazemi_nourian_nobari_movahed_2017a, keshavarzian_shamshiri_charmiyan_moaveni_2018a}. Here, we focus on the possibility to enhance mixing in micro-channels by using elastic walls: the interaction between the soft wall and the flow results in a dynamical instability, which induces transition at very low Reynolds numbers \citep{verma_kumaran_2013a}. In particular, previous linear stability studies \citep{kumaran_1996a, shankar_kumaran_1999a, kumaran_muralikrishnan_2000a} have shown that the flow over elastic walls is unstable to infinitesimal disturbances when the Reynolds number exceeds a critical value which can be tuned by decreasing the shear modulus of the soft wall, thus suggesting that there is an instability even at zero Reynolds number. The existence of this instability has been proved experimentally by \citet{verma_kumaran_2012a}, reaching a transitional Reynolds number of $200$ for the softest wall used in the experiments.

Flow instabilities at low Reynolds numbers have been previously observed in the presence of elasticity; in particular, so-called purely elastic instabilities have been reported for viscoelastic fluids in a wide variety of flow configurations and they can be generally found when inertial forces are negligible compared to elasticity  \citep{gardner_pike_miles_keller_tanaka_1982a, larson_1992a, shaqfeh_1996a, mckinley_pakdel_oztekin_1996a, haward_mckinley_shen_2016a}. Such instabilities are due to the non-linear coupling between the flow and the constitutive equation of the non-Newtonian fluid and lead to the so-called elastic turbulence \citep{groisman_steinberg_2000a, berti_boffetta_2010a, lim_ober_edd_desai_neal_bong_doyle_mckinley_toner_2014a, kawale_marques_zitha_kreutzer_rossen_boukany_2017a, steinberg_2019a}. Here, we will extend these works by considering a simple Newtonian fluid non-linearly coupled to a viscoelastic wall, and show that a self-sustained chaotic flow can be observed.

In this work, we present new Direct Numerical Simulations (DNS) of the flow over an incompressible hyper-elastic wall at Reynolds number where turbulence cannot be sustained in channels with rigid walls and show that fluid velocity fluctuations can be sustained by tuning the wall elasticity. In the fluid part of the channel, the full incompressible Navier--Stokes equations are solved, while momentum conservation and incompressibility constraint are enforced inside the solid material. In \secref{sec:formulation}, we first discuss the flow configuration and governing equations, and then present the numerical methodology used. The effects of an hyper-elastic wall on the channel flow are presented in \secref{sec:result}. Finally, a summary of the main findings and some conclusions are drawn in \secref{sec:conclusion}.

\begin{figure}
  \centering
  \includegraphics[width=0.35\textwidth]{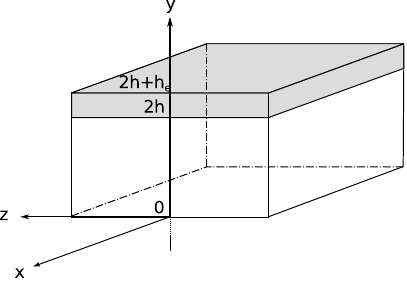}
  \caption{Sketch of the channel considered in the present work: two solid walls are located at $y=0$ and $2h+h_e$, while $y=2h$ indicates the interface between the fluid region and the elastic layer.}
  \label{fig:sketch}
\end{figure}

\section{Formulation} \label{sec:formulation}
We consider the flow of an incompressible viscous fluid through a channel with an incompressible hyper-elastic wall. A sketch of the geometry and the Cartesian coordinate system are reported in \figrefS{fig:sketch}: $x$, $y$ and $z$ denote the streamwise, wall-normal and spanwise coordinates, and $u$, $v$ and $w$ the corresponding velocity components. The channel is bounded by two rigid walls located at $y=0$ and $2h+h_e$, while the elastic layer extends from $y=2h$ to $2h+h_e$, where $h_e$ represents the height of the layer, fixed here to $h_e=0.5h$. In this work, we assume the interface of the elastic layer to be initially flat. Periodic boundary conditions are imposed in the streamwise and spanwise directions. 

The fluid and solid phase motion is governed by the conservation of momentum and the incompressibility constraint:
\begin{align}
\label{eq:NS}
\frac{\partial u_i^p}{\partial t} + \frac{\partial u_i^p u_j^p}{\partial x_j} = \frac{1}{\rho} \frac{\partial \sigma_{ij}^p}{\partial x_j} \;\;\; \textrm{and} \;\;\; \frac{\partial u_i^p}{\partial x_i} = 0,
\end{align}
where the suffix $^p$ is used to distinguish the fluid $^f$ and solid $^s$ phases. In the previous set of equations, $\rho$ is the density (assumed to be the same for the solid and fluid), and $\sigma_{ij}$ the Cauchy stress tensor. The two phases are coupled at the interface by the continuity of the velocity and traction force, \ie $u_i^f = u_i^s$ and $\sigma_{ij}^f n_j = \sigma_{ij}^s n_j$, where $n_i$ denotes the normal to the interface.

To numerically solve the fluid-structure interaction problem at hand, we introduce a monolithic velocity vector field $u_i$ valid everywhere, found by a volume averaging procedure. In particular, we introduce an additional variable $\phi^s$ which is the solid volume fraction; this is zero in the fluid and one in the solid, with $0\le\phi^s\le1$ around the interface. By doing so, we can now write the stress in a mixture form as
\begin{equation}
\label{eq:phi-stress}
\sigma_{ij} = \left( 1 - \phi^s \right) \sigma_{ij}^f + \phi^s \sigma_{ij}^s.
\end{equation}
This is the so-called one-continuum formulation \cite{tryggvason_sussman_hussaini_2007a}. The fluid is Newtonian and the solid is an incompressible viscous hyper-elastic material with constitutive equations
\begin{equation}
\label{eq:stress}
\sigma_{ij}^f = -p \delta_{ij} + 2 \mu \mathcal{D}_{ij} \;\;\;\; \textrm{and} \;\;\;\; \sigma_{ij}^s = -p \delta_{ij} + 2 \mu \mathcal{D}_{ij} + G \mathcal{B}_{ij},
\end{equation}
where $p$ is the pressure, $\mu$ the dynamic viscosity (assumed to be the same in the two phases), $\mathcal{D}_{ij}$ the strain rate tensor defined as $\mathcal{D}_{ij}=\left( \partial u_i/\partial x_j + \partial u_j/\partial x_i \right)/2$ and $\delta_{ij}$ is the Kronecker delta. The last term in the solid Cauchy stress tensor $\sigma_{ij}^s$ is the hyper-elastic contribution modelled as a neo-Hookean material, thus satisfying the incompressible Mooney-Rivlin law, where $\mathcal{B}_{ij}$ is the left Cauchy-Green deformation tensor and $G$ the modulus of transverse elasticity. The full set of equations can be closed in a purely Eulerian manner by updating $\mathcal{B}_{ij}$ and $\phi^s$ with the following transport equations:
\begin{equation}
\label{eq:adv}
\frac{\partial \mathcal{B}_{ij}}{\partial t} + \frac{\partial u_k \mathcal{B}_{ij}}{\partial x_k} = \mathcal{B}_{kj}\frac{\partial u_i}{\partial x_k} + \mathcal{B}_{ik}\frac{\partial u_j}{\partial x_k} \;\;\;\; \textrm{and} \;\;\;\; \frac{\partial \phi^s}{\partial t} + \frac{\partial u_k \phi^s}{\partial x_k} = 0.
\end{equation}

\subsection{Numerical implementation}
The previous set of equations are solved numerically: the time integration is based on an explicit fractional-step method \citep{kim_moin_1985a}, where all the terms are advanced with the third order Runge-Kutta scheme, except the solid hyper-elastic contribution which is advanced with the Crank-Nicolson scheme \citep{min_yoo_choi_2001a}. The governing differential equations are solved on a staggered grid using a second order central finite-difference scheme, except for the advection terms in \equref{eq:adv} where the fifth-order WENO scheme is applied. The code has been extensively validated, and more details on the numerical scheme and validation campaign are reported in Refs.\ \onlinecite{rosti_brandt_2017a, rosti_brandt_2018a, rosti_brandt_mitra_2018a, izbassarov_rosti_niazi-ardekani_sarabian_hormozi_brandt_tammisola_2018a, alghalibi_rosti_brandt_2019a, rosti_pramanik_brandt_mitra_2020a}; more details on the numerical method can be found in \citet{sugiyama_ii_takeuchi_takagi_matsumoto_2011a}.

For all the flows considered hereafter, the equations of motion are discretised on a fixed, Cartesian and uniform mesh with $1296 \times 540 \times 648$ grid points on a computational domain of size $6hk \times 2.5h \times 3hk$ in the streamwise, wall-normal and spanwise directions. $k$ is a factor used to increase the size of the domain in the homogeneous direction as the Reynolds number decreases \citep{tsukahara_seki_kawamura_tochio_2005a}; in particular, $k=1$, $4.30$, $18.5$, $79.5$ and $237$ for $Re_b=2800$, $651$, $151$, $35$ and $11$, respectively. The spatial resolution has been chosen in order to properly resolve the wall deformation for all the Reynolds numbers considered in the present study \citep{rosti_brandt_2017a}.

\section{Results} \label{sec:result}
We study laminar and turbulent channel flows over viscous hyper-elastic walls, together with the baseline cases over stationary impermeable walls. All the simulations are performed at constant flow rate, and thus the pressure gradient needed to drive the flow is determined at every time step to ensure this condition; it oscillates around a constant value at statistical state. The flow Reynolds number is defined based on the bulk velocity, \ie $Re_b=\rho U_b h/\mu$, where $U_b$ is the average value of the mean velocity computed across the whole domain occupied by the fluid phase; the choice of using $U_b$ and $h$ as reference velocity and length facilitates the comparison between the flow in a channel with elastic walls and the flow in a channel bounded by rigid walls. In the present work, we vary the bulk Reynolds number $Re_b$ and the modulus of transverse elasticity $G$. The full set of simulations is reported in \tabref{tab:cases}. All the simulations start with a fully developed turbulent flow over rigid walls, and then after an initial transient, a new statistically steady state solution is reached, either laminar or turbulent.

\begin{table}
\centering
\setlength{\tabcolsep}{5pt}
\begin{tabular}{ccccc|ccccc}
$Re_b$	&	$G/\left( \rho U_b^2 \right)$	&	$\overline{Re}_\tau$	&	$\overline{u}_M/U_b$		&	$\widehat{y}_M/h$	&	$Re_b$	&	$G/\left( \rho U_b^2 \right)$	&	$\overline{Re}_\tau$	&	$\overline{u}_M/U_b$		&	$\widehat{y}_M/h$			\\
\hline
$2800$	&	$\infty$									&	$180.0$		&	$1.16$											&	$~~0.000$	&	$151$		&	$\infty$									&	$21.3$		&	$1.50$										&	$~~0.000$	\\
$2800$	&	$4.0$										&	$180.8$		&	$1.17$											&	$-0.030$	&	$151$		&	$1.0$										&	$21.8$		&	$1.47$										&	$-0.069$	\\
$2800$	&	$2.0$										&	$203.3$		&	$1.19$											&	$-0.076$	&	$151$		&	$0.5$										&	$51.0$		&	$1.41$										&	$-0.222$	\\
$2800$	&	$1.0$										&	$240.5$		&	$1.23$										&	$-0.206$	&	$35$		&	$\infty$									&	$10.2$		&	$1.50$										&	$~~0.000$	\\
$2800$	&	$0.5$										&	$337.0$		&	$1.32$										&	$-0.386$	&	$35$		&	$1.0$										&	$11.2$			&	$1.49$										&	$-0.007$	\\
$651$		&	$\infty$									&	$49.8$		&	$1.29$										&	$~~0.000$	&	$35$		&	$0.5$										&	$15.3$		&	$1.48$										&	$-0.014$	\\
$651$		&	$2.0$										&	$49.8$		&	$1.30$										&	$-0.010$	&	$11$		&	$\infty$									&	$5.7$		&	$1.50$										&	$~~0.000$	\\
$651$		&	$1.0$										&	$68.3$		&	$1.33$										&	$-0.031$	&	$11$		&	$0.5$										&	$5.7$		&	$1.50$										&	$~~0.000$ \\
$651$		&	$0.5$										&	$151.1$		&	$1.37$											&	$-0.463$	&	~				&	~												&	~				&		~												&		~				
\end{tabular}
\caption{Summary of the DNSs performed, all with fixed thickness of the elastic layer $h_e=0.5h$. The table reports the bulk Reynolds number $Re_b$, the shear elastic modulus $G$, the mean friction Reynolds number $\overline{Re}_\tau$, the maximum velocity $\overline{u}_M$, and its distance from the channel centerline $\widehat{y}_M = y_M - h$.}
\label{tab:cases}
\end{table}

The friction velocity $u_\tau$ will be often employed in the following and is defined here as
\begin{equation} \label{eq:friction_velocity_total}
\overline{u}_\tau = \sqrt{\frac{\mu}{\rho} \dfrac{d \overline{u}}{d y} - \overline{u'v'} + \frac{G}{\rho} \overline{B}_{12} },
\end{equation}
where the quantities are evaluated at the mean interface location, $y=2h$. In the previous relation and in the rest of the work, the overline and the prime represent the mean and fluctuation obtained by averaging over the homogeneous directions and in time. The previous definition is used because, when the channel has moving walls, the friction velocity needs to account for the Reynolds and the elastic shear stress, that are in general non-zero at the solid-fluid interface. Note that, the actual value of the friction velocity of the elastic wall is computed from its friction coefficient, found by combining the information of the total $C_f$, obtained from the driving streamwise pressure gradient, and the one of the lower rigid wall \citep[see also][]{rosti_brandt_2017a}.

\begin{figure}
  \centering
  \includegraphics[width=0.49\textwidth]{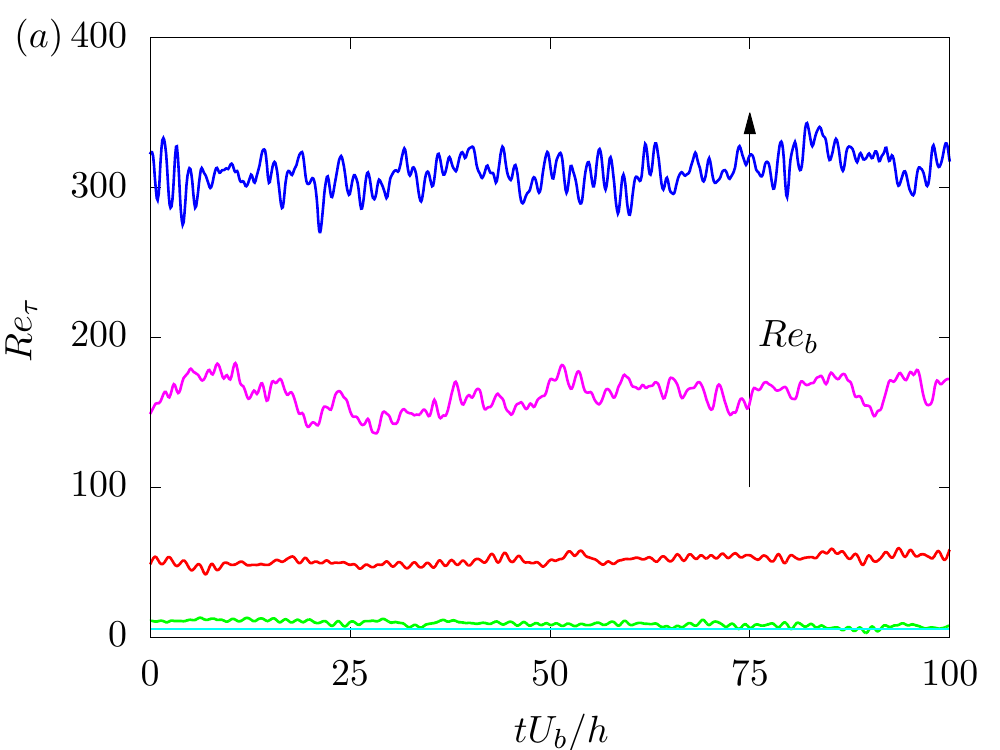}
  \includegraphics[width=0.49\textwidth]{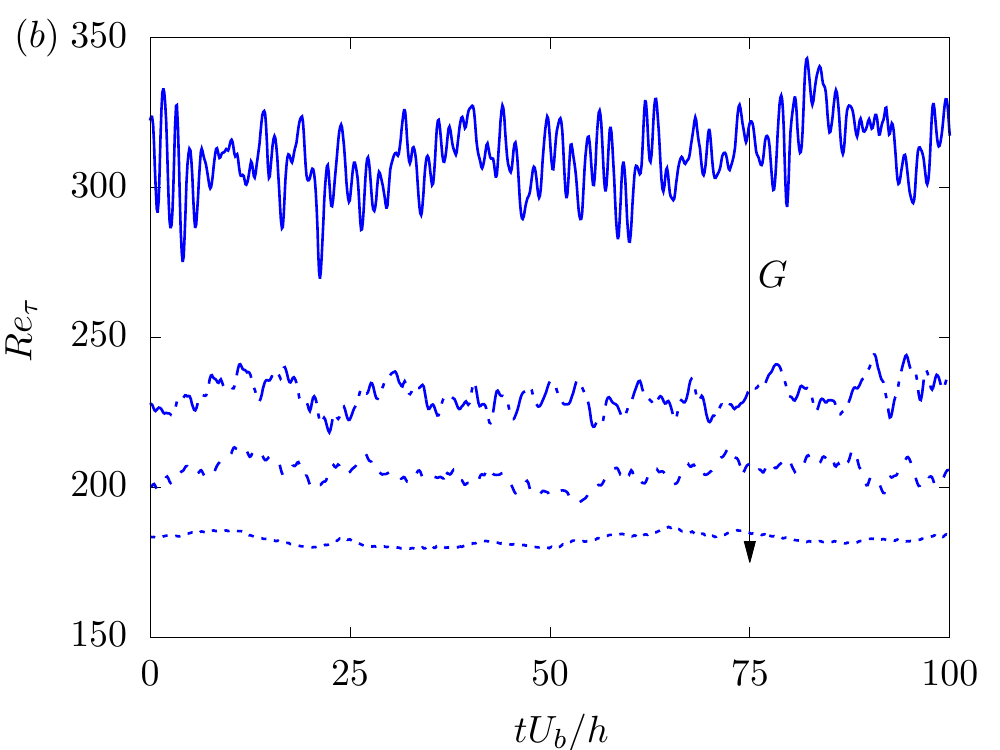}
  \caption{Time history of the friction Reynolds number $Re_\tau$ space averaged over the wall. The blue, magenta, red, green and cyan lines are used to distinguish the different bulk Reynolds numbers, $Re_b=2800$, $651$, $151$, $35$ and $11$, while the solid, dash-dotted, dashed and dotted lines to distinguish the different shear elastic moduli, $G/\rho U_b^2=0.5$, $1$, $2$ and $4$. In panel $(a)$, the amplitude of the fluctuations is amplified by a factor $5$ (magenta line), by $10$ (red line) and by $20$ (green line).}
  \label{fig:reynoldsTime}
\end{figure}
We start our analysis by studying in \figrefS{fig:reynoldsTime} the time evolution of the friction Reynolds number $Re_\tau$, \ie $u_\tau h/\nu$. In particular, panel a) shows the friction Reynolds number for the cases with the minimum elastic modulus $G=0.5 \rho U_b^2$, thus corresponding to the most deformable wall, and for different bulk Reynolds number $Re_b$. We observe that, as expected, the friction Reynolds number decreases with the bulk Reynolds number and also the amplitude of its fluctuations. However, differently from the flow over rigid walls, the flow remains unstable even for very low Reynolds numbers, $Re_b=35$ in this case, while a further reduction of the Reynolds number leads to the flow laminarisation. If we fix the bulk Reynolds number $Re_b$ and vary only the elastic modulus $G$, three different behaviors can be observed, as shown by the space and time averaged friction Reynolds number $\overline{Re}_\tau$ pertaining all cases studied in the present work collected in \figref[$a$]{fig:reynolds}: \textit{(i)} for high $Re_b$, as $G$ increases (the wall becomes more rigid) $\overline{Re}_\tau$ decreases eventually saturating at the value obtained for a turbulent flow over rigid walls, see also the time histories in \figref[$b$]{fig:reynoldsTime}; \textit{(ii)} for intermediate $Re_b$, as $G$ increases $\overline{Re}_\tau$ decreases eventually leading to a  fully laminar flow, the friction assuming the same value obtained for a laminar flow over rigid walls; \textit{(iii)} for low Reynolds numbers, the flow always becomes laminar for any initial condition and the friction Reynolds number is the same obtained for a laminar flow over rigid walls. Indeed, the thin lines in \figref[$a$]{fig:reynolds} display the characteristic values for laminar and turbulent channel flows. For every $Re_b$, reducing the wall elasticity implies a reduction of the resulting $\overline{Re}_\tau$; all cases converge to the rigid wall solution as $G$ increases, in particular, $\overline{Re}_\tau$ converges to the turbulent experimental correlation $0.09 \left( 2Re_b \right)^{0.88}$ (see \eg Ref.~\onlinecite{pope_2001a}) for $Re_b \gtrsim 482$ and to the laminar analytical solution $\sqrt{3Re_b}$ for $Re_b \lesssim 482$ as $G \rightarrow \infty$.

\begin{figure}
  \centering
  \includegraphics[width=0.49\textwidth]{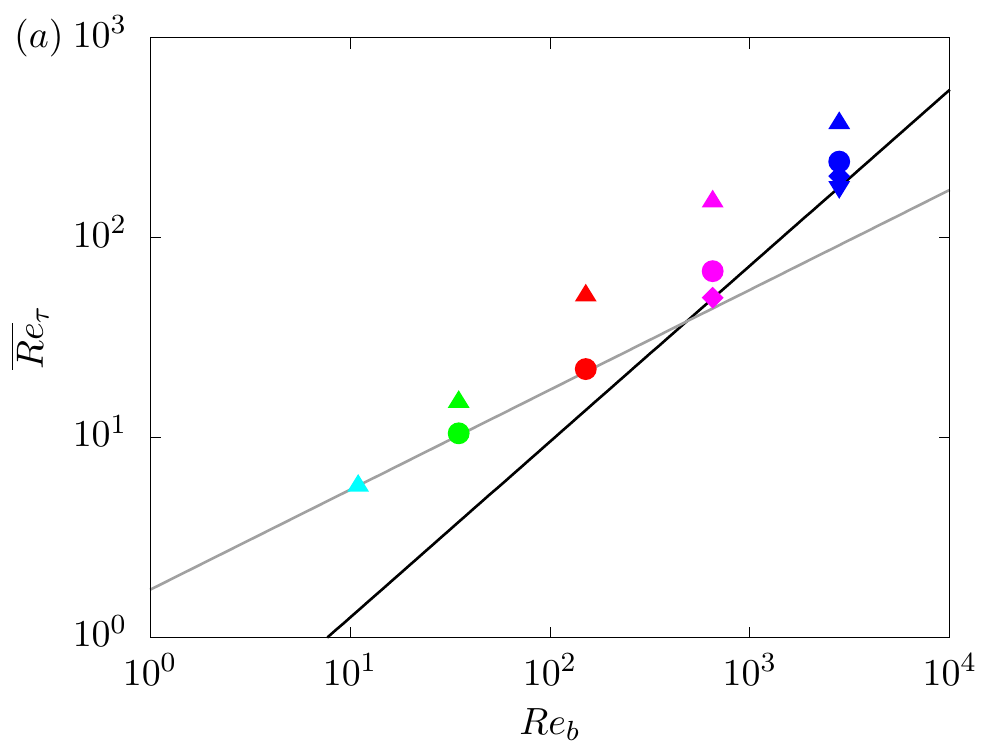}
  \includegraphics[width=0.49\textwidth]{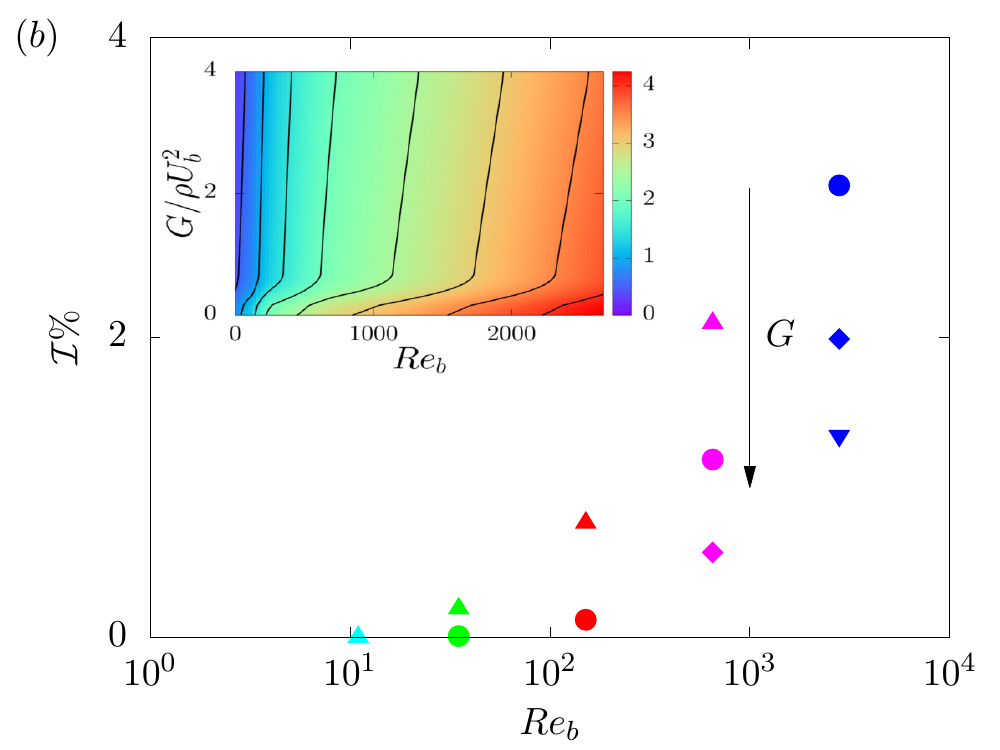}
  \caption{$(a)$ Mean friction Reynolds number $\overline{Re}_\tau$ and $(b)$ root mean square of the friction velocity normalised by its mean, \ie $\mathcal{I}=\sqrt{\overline{u_\tau' u_\tau'}}/\overline{u}_\tau$, as a function of the bulk Reynolds number. The blue, magenta, red, green and cyan colors are used to distinguish different bulk Reynolds numbers $Re_b=2800$, $651$, $151$, $35$ and $11$, while the upper-triangle $\blacktriangle$, circle {\Large $\bullet$}, rombus $\blacklozenge$ and lower-triangle $\blacktriangledown$ to distinguish different shear elastic moduli $G/\rho U_b^2=0.5$, $1$, $2$ and $4$. The grey and black lines in panel a) are the analytical solutions for laminar flows and the experimental correlation for turbulent flows, respectively. The inset in panel b) reports the contour of $\mathcal{I}\%$ as a function of the Reynolds number $Re_b$ and elastic shear modulus $G$ obtained by interpolation and extrapolation of our data. The black lines are separated by $0.5$.}
  \label{fig:reynolds}
\end{figure}

\begin{figure}
  \centering
  \includegraphics[width=0.49\textwidth]{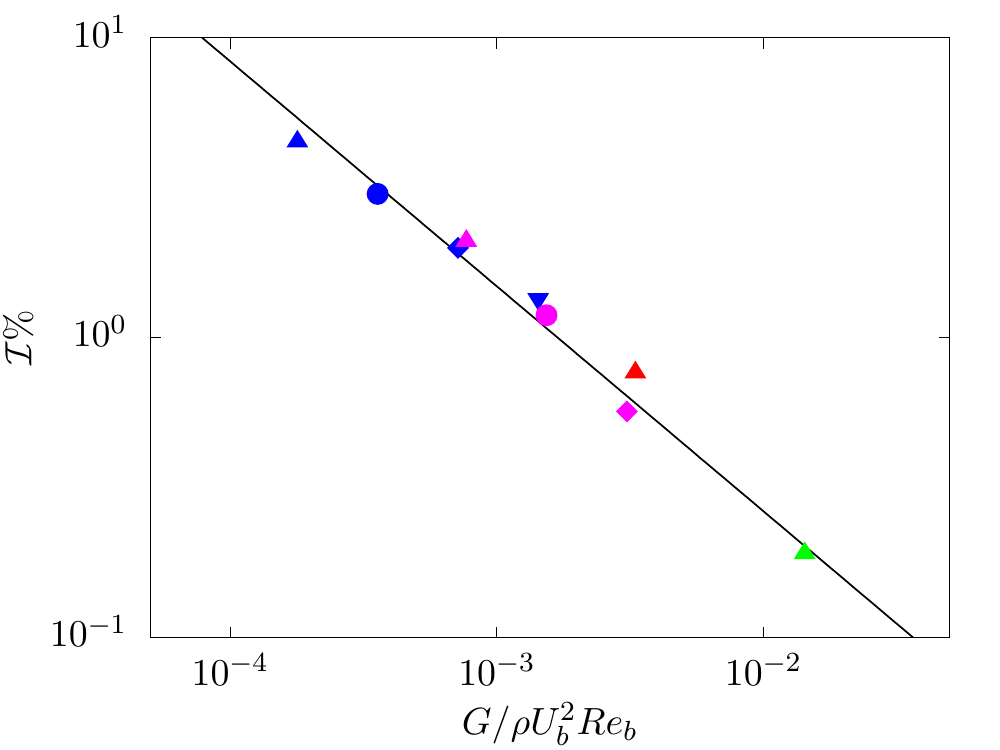}
  \caption{Root mean square of the friction velocity normalised by its mean, \ie $\mathcal{I}=\sqrt{\overline{u_\tau' u_\tau'}}/\overline{u}_\tau$, as a function of the ratio of the normlaised wall elasticity $G$ and bulk Reynolds number, \ie $G/\rho U_b^2 Re_b$. The blue, magenta, red and green colors are used to distinguish different bulk Reynolds numbers $Re_b=2800$, $651$, $151$ and $35$, while the upper-triangle $\blacktriangle$, circle {\Large $\bullet$}, rombus $\blacklozenge$ and lower-triangle $\blacktriangledown$ to distinguish different shear elastic moduli $G/\rho U_b^2=0.5$, $1$, $2$ and $4$.}
  \label{fig:elasticity}
\end{figure}
To quantify the unsteady nature of the flow, we compute the root mean square of the friction velocity $\sqrt{\overline{u_\tau' u_\tau'}}$, used here as a measure of the flow fluctuations. This is divided by its mean value and reported in \figref[$b$]{fig:reynolds} as a function of the bulk Reynolds number for all the cases considered here. Consistently with the previous discussion, we observe that reducing the wall elasticity induces a reduction of the fluctuations. This reduction is strongly non-linear, with large reduction for increment in small values of $G$ and small reduction for increment in large values of $G$. Also, we can observe again that the high Reynolds number cases converge, as $G$ increases, to a non-zero level of fluctuations, \ie the turbulent rigid wall solution, while the low Reynolds number cases tend to the laminar solution with zero fluctuations. Reducing the Reynolds number, we observe a further reduction of the fluctuation intensity; also in this case the variation is strongly non-linear with large reductions of the fluctuation intensity for large Reynolds numbers, while smaller variation are observed at small $Re_b$, when the flow tends to become laminar. The inset of \figref[$b$]{fig:reynolds} shows the same quantity, $\mathcal{I}$, as a function of both $Re_b$ and $G$ as a contour plot obtained by interpolating and extrapolating our data. We observe that, although in general $\mathcal{I}$ is a function of both $Re_b$ and $G$, \ie $\mathcal{I} = \mathcal{F} \left( Re_b, G \right)$, there is a critical value $G^* \left( Re_b\right)$ above which the solution does not significantly change anymore with the wall elasticity, and thus $\mathcal{I} = \mathcal{F}_r \left( Re_b \right)$ for $G>G^*$, where $\mathcal{F}_r$ is the solution for the flows over rigid walls. On the other hand, for $G<G^*$ the solution strongly depends on the wall elasticity: this suggests that it is possible to maintain an unsteady chaotic turbulent-like flow in principle for any Reynolds number down to $0$, as long as the wall shear elastic modulus $G$ is reduced accordingly. If we now replot the data in \figref[$b$]{fig:reynolds} as a function of a new quantity, obtained as the ratio of the wall elasticity $G/\rho U_b^2$ and the bulk Reynolds number $Re_b$, we obtain \figrefS{fig:elasticity}. By doing so, all the non-laminar cases successfully collapse onto a single master curve, decaying with power $-0.75$, \ie $\mathcal{I} \sim \left( G/\rho U_b^2 Re_b \right)^{-0.75}$. This behaviour further corroborates the idea that the level of fluctuations in the channel can be amplified either by increasing the Reynolds number (at fixed elasticity) or by increasing the wall flexibility, \ie reducing $G$ (at fixed Reynolds number).

\begin{figure}
  \centering
  \includegraphics[width=0.49\textwidth]{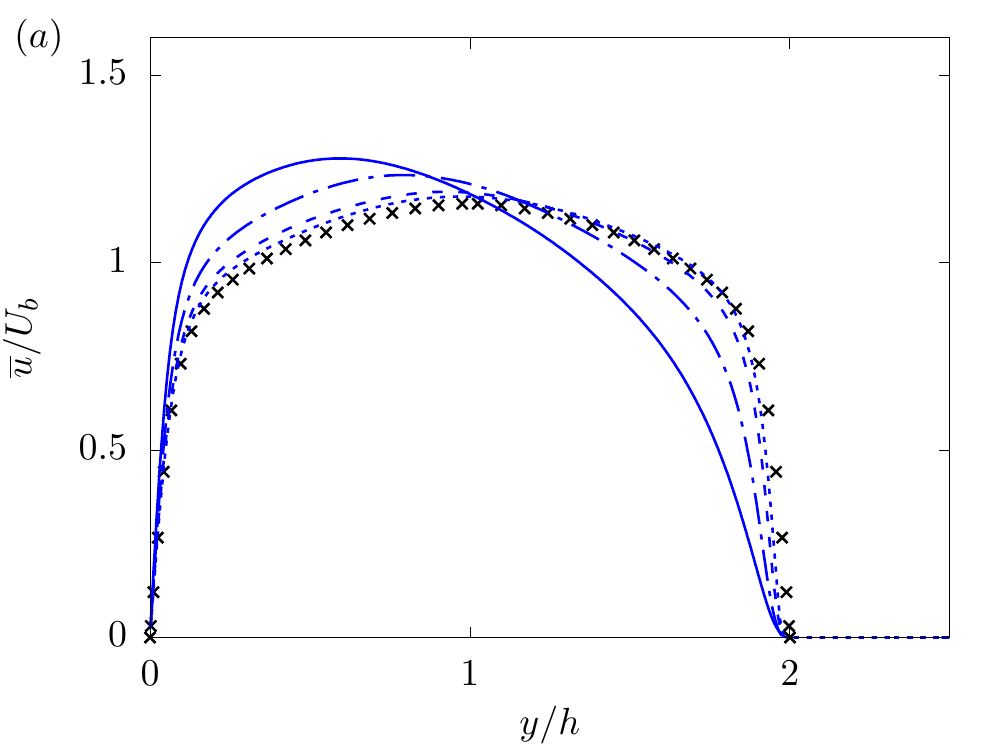}
  \includegraphics[width=0.49\textwidth]{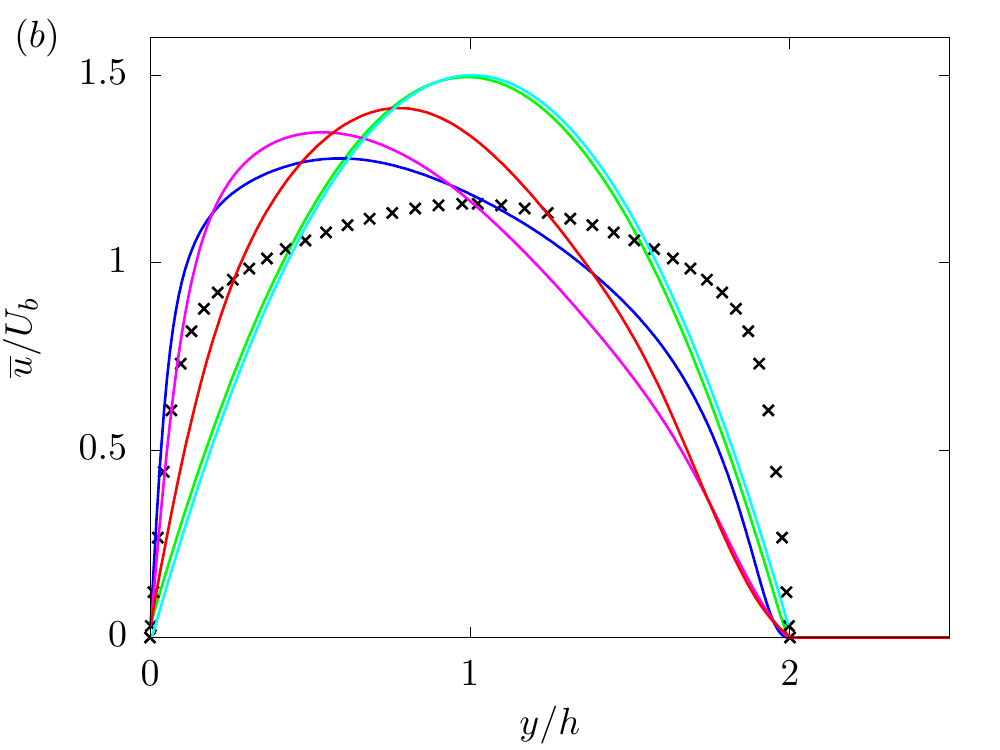}
  \caption{Mean velocity profile $\overline{u}$ as a function of the wall-normal distance $y$ for $(a)$ different wall elastic moduli $G$ at $Re_b=2800$ and for $(b)$ different Reynolds numbers $Re_b$ with $G=0.5 \rho U_b^2$. The line colors and styles are the same as in \figrefS{fig:reynoldsTime}. The symbols represent the profiles from the DNS by Kim, Moin and Moser \cite{kim_moin_moser_1987a} of turbulent flow between two solid rigid walls plotted as a reference.}
  \label{fig:U}
\end{figure}
\begin{figure}
  \centering
  \includegraphics[width=0.49\textwidth]{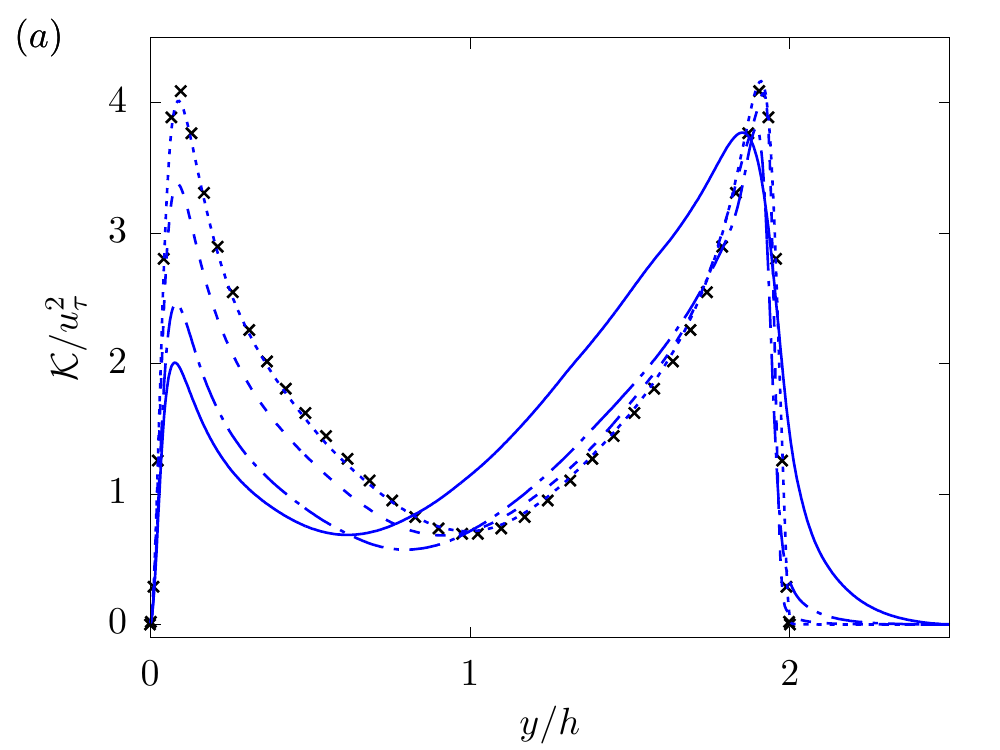}
  \includegraphics[width=0.49\textwidth]{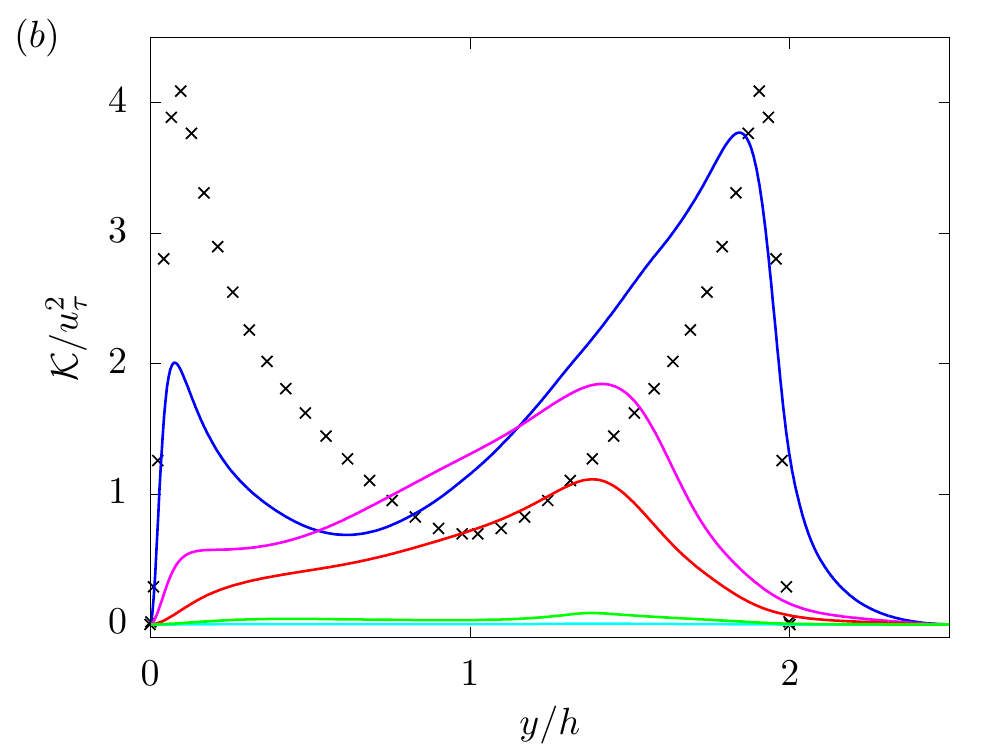}
  \caption{Mean turbulent kinetic energy $\mathcal{K}$ as a function of the wall-normal distance $y$ for $(a)$ different wall elastic moduli $G$ at $Re_b=2800$ and for $(b)$ different Reynolds numbers $Re_b$ with $G=0.5 \rho U_b^2$. The line colors and styles are the same as in \figrefS{fig:reynoldsTime}. The symbols represent the profiles from the DNS by Kim, Moin and Moser \cite{kim_moin_moser_1987a} of turbulent flow between two solid rigid walls plotted as a reference.}
  \label{fig:K}
\end{figure}

Next, we characterize the unsteady flows in terms of mean and fluctuation velocities. We start by considering the wall-normal profiles of the mean velocity $\overline{u}$ and turbulent kinetic energy $\mathcal{K} = \rho \overline{u'_i u'_i} /2$, reported in \figrefS{fig:U} and \figrefS{fig:K}. In particular, the left panels of the two figures show $\overline{u}$ and $\mathcal{K}$ at a fixed Reynolds number ($Re_b=2800$) and for all the wall elasticities $G$ studied in this work, while the right ones report $\overline{u}$ and $\mathcal{K}$ for a fixed wall elasticity ($G=0.5\rho U_b^2$) and for all the Reynolds numbers $Re_b$. From \figrefS{fig:U}, we observe that the mean velocity of the elastic wall is equal to zero \citep{rosti_brandt_2017a}; indeed, the elastic layer can only oscillates around its equilibrium position being attached to the top stationary rigid wall. Although the mean velocity is zero inside this layer, the elastic layer induces profound modification of the fluid flow in the channel. In particular, the mean velocity profile becomes more skewed, with its maximum $\overline{u}_M$ increasing and located closer to the rigid wall as the elasticity increases ($G$ decreases) as shown in \figref[$a$]{fig:U}. Note that, an inflection point in the mean profile appears within the fluid region ($0<y<2h$), usually associated to the occurrence of a Kelvin-Helmholtz instability and the formation of large scale spanwise-correlated rollers \citep{jimenez_uhlmann_pinelli_kawahara_2001a, rosti_cortelezzi_quadrio_2015a, kuwata_suga_2016a, rosti_brandt_2017a, rosti_brandt_pinelli_2018a, kuwata_suga_2019a, monti_omidyeganeh_pinelli_2019a, monti_omidyeganeh_eckhardt_pinelli_2020a}. When the Reynolds number is decreased (right panel), the asymmetry in the flow reduces with the maximum velocity increasing and its location moving back towards the channel center. Eventually, the laminar analytical profile is recovered for the smallest $Re_b$ considered.

When focusing on the velocity fluctuations in \figref[$a$]{fig:K}, we observe that the turbulent kinetic energy $\mathcal{K}$ is higher close to the elastic wall than close to the rigid wall, with the maximum value becoming almost the double of the peak close to the bottom wall for the most deformable case (left panel). This is due to the movement of the deformable wall which strongly increases the velocity fluctuations, especially the ones in the wall-normal directions, \ie $v'$ (see \eg Ref.~\onlinecite{rosti_brandt_2017a}). Furthermore, the near-wall peaks of the turbulent kinetic energy move farther from the elastic walls as the elasticity is increased. The turbulent fluctuations have non-zero values at $y = 2h$ for the elastic cases, since the no-slip condition is now enforced on a wall which is moving, \ie $u_i^f=u_i^s$. In particular, $\mathcal{K}$ does not clearly vanish until reaching the rigid top wall ($y=2.5h$), thus indicating that the fluctuations propagate deeply inside the solid layer. The asymmetry in the flow originates from the asymmetry of the geometry; this induces the shift of the minimum of $\mathcal{K}$ towards the rigid walls, as well as the shift in the same direction of the maximum velocity, as reported in \tabref{tab:cases}. 

\figrefC[$b$]{fig:K} shows how the turbulent kinetic energy $\mathcal{K}$ scales with the Reynolds number, for a fixed wall elasticity; in particular, the softest wall is considered here. We observe that, as $Re_b$ decreases the peak of turbulent kinetic energy close to the rigid wall rapidly vanishes, as expected for flows over rigid walls, where the lowest Reynolds number able to sustain a turbulent flow is around $600$, as reported by \citet{tsukahara_seki_kawamura_tochio_2005a}. A similar trend is evident for the peak close to the moving wall, but the decrease is much lower than for a rigid wall. Indeed, for $Re_b<600$ the near-wall peak close to the rigid wall completely disappear, and the profiles exhibit a single peak close to elastic wall. Also, the peak moves away from the deformable wall towards the bulk of the channel as the Reynolds number reduces, indicating that all the turbulent fluctuations in the channel at low $Re_b$ are produced by the moving wall, then propagating across the channel.

\begin{figure}
  \centering
  \includegraphics[width=0.49\textwidth]{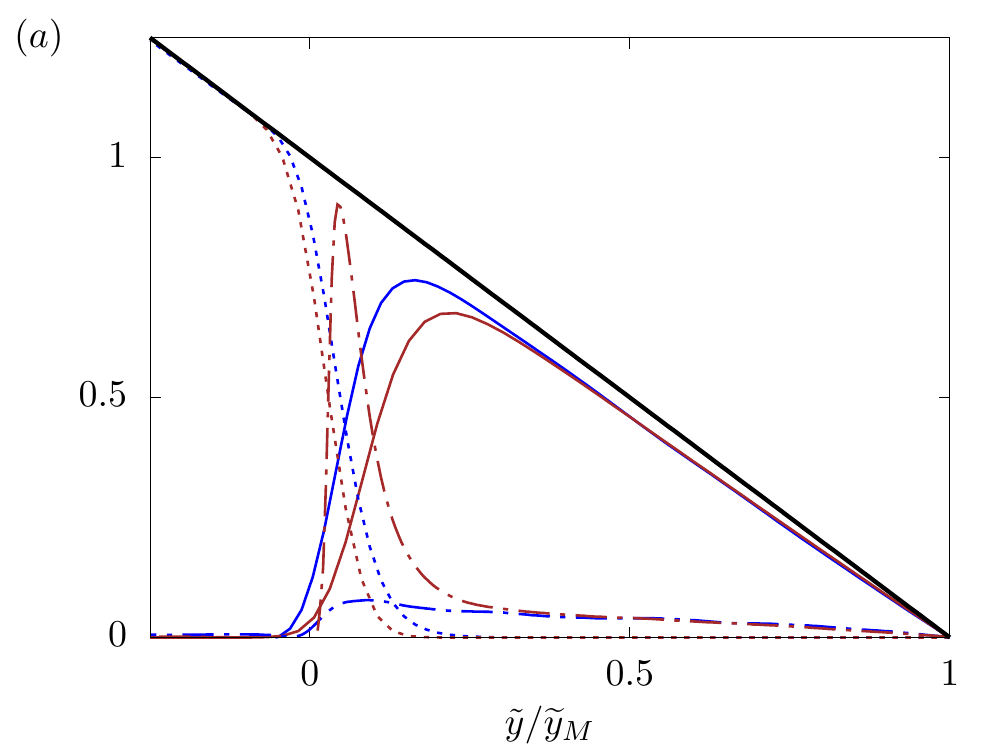}
  \includegraphics[width=0.49\textwidth]{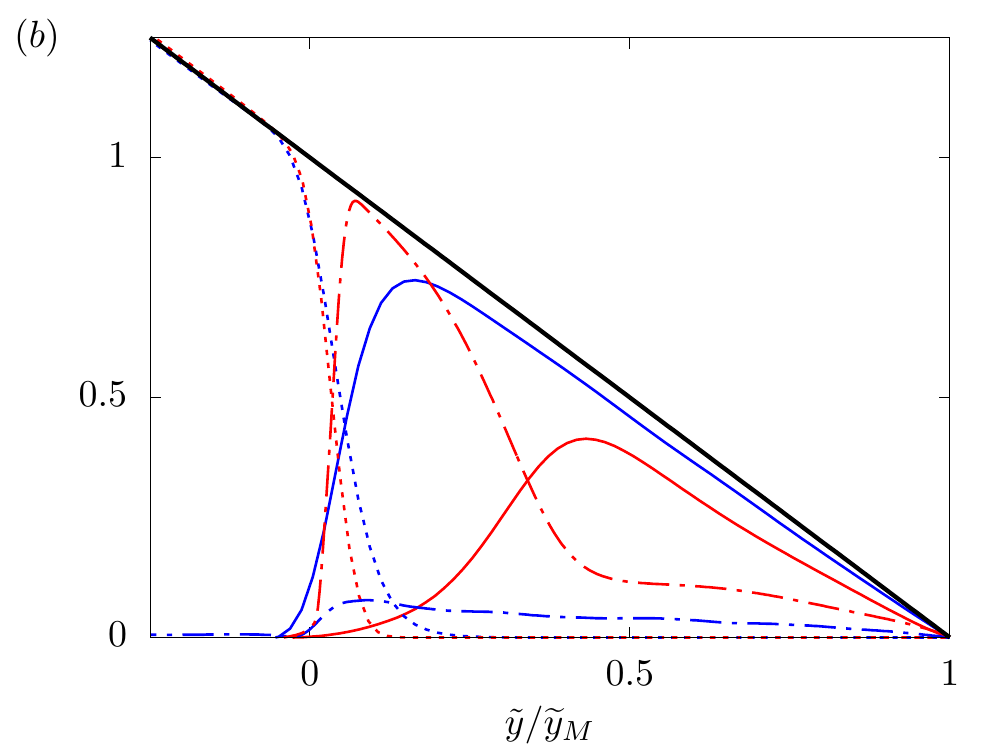}
  \caption{Reynolds shear stress $-\rho \overline{u'v'}$ (solid line), viscous stress $\mu d\overline{u}/dy$ (dash-dotted line) and shear elastic stress $G\overline{B}_{12}$(dotted line) in wall units as a function of the wall-normal distance from the elastic wall $\widetilde{y}=2h-y$ normalised by $\widetilde{y}_M=2h-y_M$ for $(a)$ different wall elastic moduli $G$ at $Re_b=2800$ and for $(b)$ different Reynolds numbers $Re_b$ with $G=0.5 \rho U_b^2$. The values of $y_M$ used to normalize the different abscissa can be obtained from \tabref{tab:cases}. In particular, the blue, brown and red colors in the two panels are used for the cases $Re_b=2800$ and $G=0.5 \rho U_b^2$, $Re_b=2800$ and $G=4 \rho U_b^2$, and $Re_b=151$ and $G=0.5 \rho U_b^2$.}
  \label{fig:stressBal}
\end{figure}
Apart from the diagonal components of the Reynolds stress tensor discussed above in terms of the turbulent kinetic energy, another important observable is the off-diagonal shear component of the Reynolds stress tensor $- \rho \overline{u'v'}$, which together with the mean viscous $\mu d\overline{u}/dy$ and elastic stress $G\overline{B}_{12}$ shear components form the total shear stress, \ie
\begin{equation} \label{eq:shearBalance}
\tau = \mu d\overline{u}/dy - \rho \overline{u'v'} + G\overline{B}_{12}.
\end{equation}
All of these are reported in \figref[$a$]{fig:stressBal} for the cases at $Re_b=2800$ (solid lines). The cross Reynolds stress component is strongly affected by the presence of the moving wall: the maximum value increases and moves away from the wall as the elasticity increases at a fixed Reynolds number. The stress profiles vary linearly in the bulk of the channel away from the wall, although with different slopes depending on $Re_b$ and $G$. Most of these effects are well compensated in the figure by dividing $\widetilde{y}$ with $\widetilde{y}_M=2h-y_M$, \ie the distance of the location of the maximum mean velocity from the elastic wall. At the interface the value of the stress is not null as in the rigid case, however, inside the elastic layer the Reynolds shear stress vanishes quickly. The mean viscous stress is almost null in the solid and in the bulk of the channel and exhibits a small peak close to the interface which increase as $G$ increases, \ie the wall is more rigid, eventually having the maximum at the interface for the completely rigid case; the elastic stress, on the contrary, is null in the fluid region and almost the total stress in the solid layer. Thus, we can conclude that the total shear stress is dominated by the elastic stress in the solid layer, by the Reynolds stress in the bulk of the channel and by the balance of all the three components at the interface, with the relative contributions at the interface strongly changing with $G$: for rigid walls the dominant and only contribution not null at the interface is the viscous stress, while for flexible walls the Reynolds and elastic stresses grow with the wall elasticity. When the Reynolds number is varied, the balance between the three terms is significantly altered, as shown in \figref[$b$]{fig:stressBal}. Indeed, as the Reynolds numbers decreases, the Reynolds shear stress peak shifts away from the wall, thus reducing its total contribution. On the other hand, the viscous contribution increases and compensates for the loss of Reynolds shear stress. For the lowest Reynolds number (not shown in the figure), the flow is fully laminar, and the total stress is equal to the elastic stress in the solid layer and to the viscous stress in the fluid region, with the Reynolds shear stress being null. From the figure we can conclude that, differently from the flow over rigid walls, the turbulent fluctuations do not rapidly vanish when reducing the Reynolds number because of their persistence in the bulk of the channel.

To confirm these observations, we consider the turbulent kinetic energy balance. To do so, we decompose the velocity field $u_i \left( x, y, z, t \right)$ into its mean $\overline{u}_i \left( y \right)$ and fluctuation $u'_i \left( x, y, z, t \right)$ as $u_i = \overline{u}_i + u'_i$. By substituting this into the governing equation, we obtain
\begin{equation}
\rho \left( \frac{\partial u'_i}{\partial t} + \frac{\partial u'_i u'_j}{\partial x_j} + \frac{\partial \overline{u}_i u'_j}{\partial x_j} + \frac{\partial u'_i \overline{u}_j}{\partial x_j} + \frac{\partial \overline{u}_i \overline{u}_j}{\partial x_j} \right)  = - \frac{\partial p}{\partial x_i} + 2\mu \frac{\partial  \mathcal{D}_{ij}}{\partial x_j} + G \frac{\partial \phi^s \mathcal{B}_{ij}}{\partial x_j},
\end{equation}
which can be rewritten for later convenience as
\begin{equation}
\rho \left( \frac{\partial u'_i}{\partial t} + \frac{\partial u'_i u'_j}{\partial x_j} + u'_j \frac{\partial \overline{u}_i}{\partial x_j} + \frac{\partial u'_i \overline{u}_j}{\partial x_j} + \frac{\partial \overline{u}_i \overline{u}_j}{\partial x_j} \right)  = - \frac{\partial p}{\partial x_i} + 2\mu \frac{\partial  \mathcal{D}_{ij}}{\partial x_j} + G \frac{\partial \phi^s \mathcal{B}_{ij}}{\partial x_j}.
\end{equation}
We now multiply the equation by $u'_i$ and obtain
\begin{equation}
\begin{split}
\rho \left( \frac{\partial u'_i u'_i/2}{\partial t} + \frac{\partial u'_i u'_i u'_j/2}{\partial x_j} + u'_i u'_j \frac{\partial \overline{u}_i}{\partial x_j} + \frac{\partial u'_i u'_i \overline{u}_j/2}{\partial x_j} + u_i \frac{\partial \overline{u}_i \overline{u}_j}{\partial x_j} \right)  = \\
- \frac{\partial u'_i p}{\partial x_i} + 2\mu \frac{\partial  u'_i \mathcal{D}_{ij}}{\partial x_j} - 2 \mu \mathcal{D}_{ij} \mathcal{D}_{ij} + G \frac{\partial u'_i \phi^s \mathcal{B}_{ij}}{\partial x_j} - G \phi^s \mathcal{B}_{ij} \mathcal{D}_{ij},
\end{split}
\end{equation}
where we made use of 
\begin{equation}
u'_i \frac{\partial \mathcal{D}_{ij}}{\partial x_j} = \frac{\partial u'_i \mathcal{D}_{ij}}{ \partial x_j} - \mathcal{D}_{ij} \frac{\partial u'_i }{ \partial x_j} = \frac{\partial u'_i \mathcal{D}_{ij}}{ \partial x_j} - \mathcal{D}_{ij} \mathcal{D}'_{ij},
\end{equation}
and similarly of
\begin{equation}
u'_i \frac{\partial \phi^s \mathcal{B}_{ij}}{\partial x_j} = \frac{\partial u'_i \phi^s \mathcal{B}_{ij}}{ \partial x_j} - \phi^s \mathcal{B}_{ij} \frac{\partial u'_i }{ \partial x_j} = \frac{\partial u'_i \phi^s \mathcal{B}_{ij}}{ \partial x_j} - \phi^s \mathcal{B}_{ij} \mathcal{D}'_{ij},
\end{equation}
where the last substitution is possible being $\mathcal{D}_{ij}$ and $\mathcal{B}_{ij}$ symmetric tensors. The equation above can then be volume averaged with the operator
\begin{equation}
\langle \cdot \rangle = \frac{1}{\mathcal{V}} \int_\mathcal{V} \cdot \ \mathrm{d} \mathcal{V},
\end{equation}
leading to the equation
\begin{equation}
\rho \left( \frac{\partial \langle u'_i u'_i \rangle/2}{\partial t} + \langle u'_i u'_j \rangle \frac{\partial \overline{u}_i}{\partial x_j} \right)  = - 2 \mu \langle \mathcal{D}'_{ij} \mathcal{D}'_{ij} \rangle - G \langle \phi^s \mathcal{B}_{ij} \mathcal{D}'_{ij} \rangle.
\end{equation}
Here, all the transport terms $\langle \partial u'_i \mathcal{F}_{ij}/\partial x_j \rangle$ vanish due to the homogeneity of the domain and to the no-slip and no-penetration boundary conditions at the rigid walls, and the terms $\langle u'_i \partial \overline{u}_i \overline{u}_j / \partial x_j \rangle$ and $\langle \overline{\mathcal{D}}_{ij} \mathcal{D}'_{ij} \rangle$ because $\langle u'_i \rangle = 0$ and $\langle \mathcal{D}'_{ij} \rangle = 0$ due to ergodicity. Finally, we obtain the turbulent kinetic energy equation
\begin{equation}
\label{eq:tke}
\frac{d \mathcal{K}}{d t} = \mathcal{P} - \varepsilon - \psi_G,
\end{equation}
where the different terms indicate the rate of change of turbulent kinetic energy $\mathcal{K}$, the turbulent production rate $\mathcal{P}$, the dissipation rate $\varepsilon$ and the power of the elastic wall $\psi_G$, defined as
\begin{equation}
\mathcal{K} = \rho \langle u'_i u'_i \rangle/2, \ \
\mathcal{P} = -\rho \langle u'_1 u'_2 \frac{\partial \overline{u}_1}{\partial x_2} \rangle, \ \
\varepsilon = 2 \mu \langle \mathcal{D}'_{ij} \mathcal{D}'_{ij} \rangle, \ \
\psi_G = G \langle \phi^s \mathcal{B}_{ij} \mathcal{D}'_{ij} \rangle.
\end{equation}
$\psi_G$ is the rate of work performed by the fluid on the elastic wall and can be either positive or negative and thus a sink or source of turbulent kinetic energy. At statistically steady state, the time derivative is obviously null, and thus \equref{eq:tke} reduces to a balance between $\mathcal{P}$, $\varepsilon$ and $\psi_G$.

\begin{figure}
  \centering
  \includegraphics[width=0.49\textwidth]{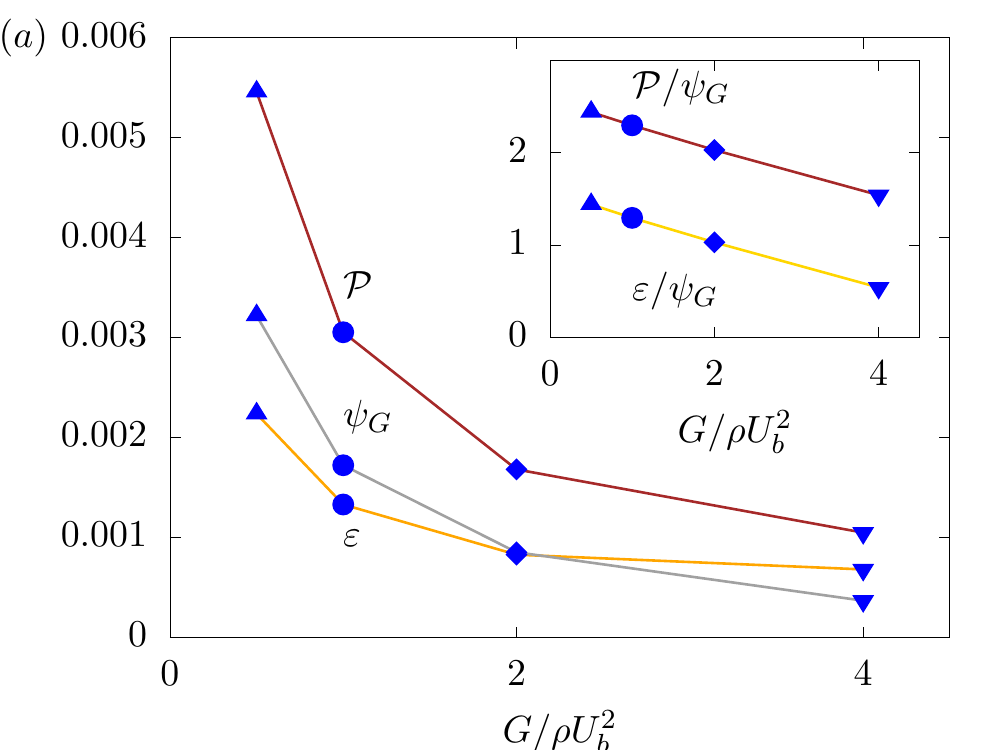}
  \includegraphics[width=0.49\textwidth]{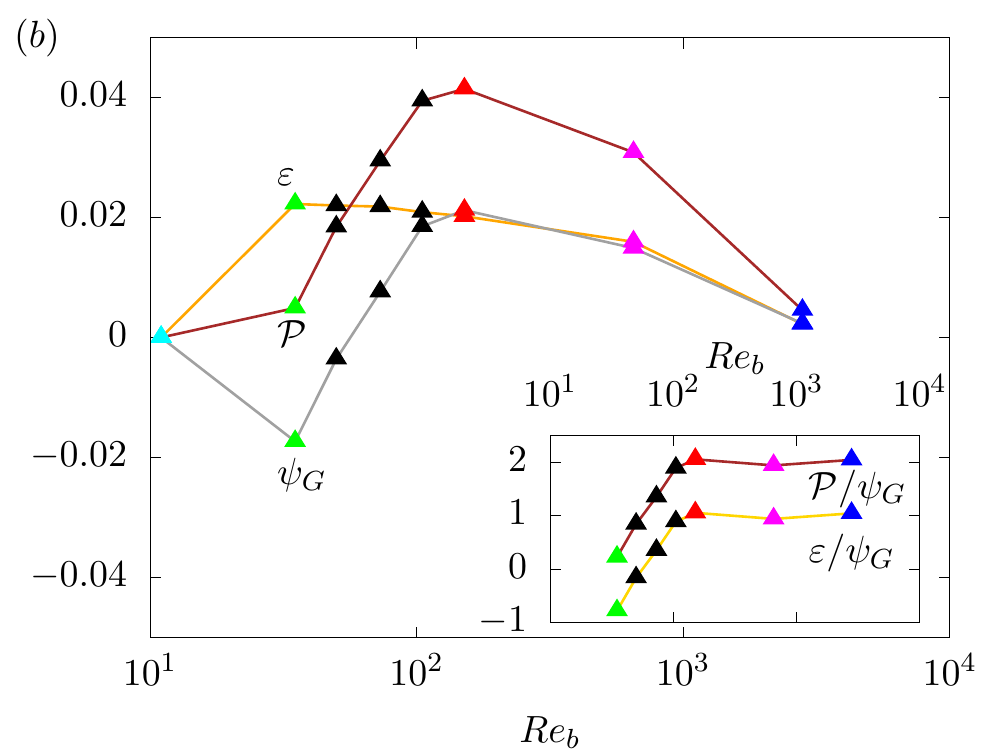}
  \caption{Volume averaged turbulent production $\mathcal{P}$ (brown), turbulent dissipation $\varepsilon$ (orange) and power of the elastic wall $\psi_G$ (grey) as a function of the wall elasticity $G$ for a fixed Reynolds number $Re_b=2800$ $(a)$ and as a function of the Reynolds number $Re_b$ for a fixed wall elasticity $G=0.5 \rho U_b^2$ $(b)$. The symbol style is the same as in \figrefS{fig:reynolds} with the addition of the black triangles,  additional simulations included for the sake of clarity. The two inset figures show the production $\mathcal{P}$ and dissipation $\varepsilon$ rates divided by the power of the elastic wall $\psi_G$.}
  \label{fig:kinBal}
\end{figure}

These three terms are displayed in \figrefSC{fig:kinBal}  as a function of the shear elastic modulus $G$ (left panel) and of the Reynolds number $Re_b$ (right panel). In the left panel we see that the elastic power contribution is positive, and indeed the presence of the elastic wall acts as an additional dissipation term at high Reynolds number. This term reduces as $G$ increases, eventually vanishing for perfectly rigid walls when $G \rightarrow \infty$. On the other hand, the behavior at fixed $G$ is non-monotonic with $Re_b$: as $Re_b$ decreases all the terms first increase, reach a maximum and then decreases. In particular, all the terms grow by a factor of around $10$ when decreasing the Reynolds number from $2800$ to $151$. Interestingly, while the turbulent production rapidly vanishes as the flow is approaching the laminar flow (for $Re_b\lesssim 151$), the power of the elastic walls change sign and becomes a production term for the turbulent kinetic energy. Because of this, the flow can remain turbulent at much lower Reynolds numbers than what usually found for flows over rigid walls and, by choosing properly the value of $G$, fluctuations can be sustained at any small Reynolds number. In conclusion, while at high Reynolds number the standard wall cycle \citep{jimenez_pinelli_1999a} takes place (although slightly modified by the elastic walls \citep{rosti_brandt_2017a}), at low Reynolds number a different mechanism arises to sustain the chaotic flow: this new mechanism originates from the non-linear interaction between the elastic solid and the fluid and resembles what found at low Reynolds and high Weissenberg numbers (\ie high elasticity numbers) for non-Newtonian fluids \citep{groisman_steinberg_2000a, haward_mckinley_shen_2016a, hopkins_haward_shen_2020a}.

\section{Conclusions} \label{sec:conclusion}
We have carried out a number of direct numerical simulations of laminar and turbulent channel flows over a viscous hyper-elastic wall. The flow inside the fluid region is described by the Navier--Stokes equations, while momentum conservation and incompressibility are imposed inside the solid layer. The two sets of equations are coupled using a one-continuum formulation allowing a fully Eulerian description of the multiphase flow problem. Here, we systematically reduce  the Reynolds number and vary the wall elasticity to identify in which condition a chaotic unsteady flow can be sustained.

In general, the friction Reynolds number $Re_\tau$ is a function of both the bulk Reynolds number $Re_b$ and the wall shear elastic modulus $G$: we show that, reducing the the wall elasticity leads to a reduction of the resulting friction Reynolds number, with the value converging to the value of the turbulent flow over rigid walls for $Re_b\gtrsim482$ and to the laminar analytical solution for $Re_b\lesssim482$. There is therefore a critical value $G^*$ above which the solution does not change anymore with the wall elasticity and the flow behaves as in the presence of rigid walls. More interestingly, for $G<G^*$ the solution depends on the wall elasticity: the mean friction and the velocity fluctuations increase with the wall deformability and it is possible to maintain an unsteady chaotic turbulent-like flow in principle for any Reynolds number, \ie in conditions where a standard flow over rigid walls would be laminar, as long as the wall shear elastic modulus $G$ is properly reduced.

We show that, at low Reynolds number, the velocity fluctuations are mainly generated by the elastic wall, while the fluctuations close to the rigid wall rapidly vanish. As we reduce $Re_b$ to values of order $100$, we observe an increase of the velocity fluctuations due to strong wall oscillations, associated to an increase of the turbulent production $\mathcal{P}$. The power of the elastic wall is a dissipation term, approximately of the same order of the viscous dissipation, thus promoting the fragmentation of typical coherent structures and the consequent formation of small scale structures. Further reducing the bulk Reynolds number, $\mathcal{P}$ decreases as the Reynolds stresses decrease in the shear layer close to the elastic wall and remain strong only in the bulk of the channel where the mean shear is negligible. On the other hand, the power of the elastic wall changes sign and becomes a source of turbulence kinetic energy, mostly balanced by the viscous dissipation. At fixed shear elastic modulus, the flow eventually laminarises, which can be compensated by a reduction of $G$, which monotonically increases the fluctuations in the flow. Indeed, we found that the level of fluctuations scale approximately as $\sim \left( G/Re_b \right)^{-0.75}$. Thus, we can conclude that the chaotic flow at very low Reynolds numbers is mainly sustained by the elastic wall oscillations, which produce turbulent kinetic energy at the interface, then transferred to the fluid through viscous stresses; this process sustains non zero Reynolds stresses in the bulk of the channel.

The present results can have profound influence on the development of strategies to increase mixing in microfluidic devices by exploiting a dynamical instability associated to the coupling between the flow and an elastic wall.

\section*{Acknowledgments}
The authors acknowledge computer time provided by the Swedish National Infrastructure for Computing (SNIC ) and by the Scientific Computing section of Research Support Division at OIST. L.B. acknowledges financial support by the Swedish Research Council, VR 2016-06119, Hybrid multiscale modelling of transport phenomena for energy efficient processes.

\section*{Data Availability Statement}
The data that support the findings of this study are available from the corresponding author upon reasonable request.


\end{document}